\documentclass[conference]{IEEEtran}

\usepackage{cite}

\ifCLASSINFOpdf
  \usepackage[pdftex]{graphicx}
  
  \DeclareGraphicsExtensions{.pdf,.jpeg,.png}
\else
 
\fi

\usepackage[hidelinks]{hyperref}
\usepackage{url}
\usepackage{svg}
\usepackage{amsmath,amssymb,amsfonts}
\usepackage{algorithmic}
\usepackage{graphicx}
\usepackage{textcomp}
\usepackage{booktabs}
\usepackage{siunitx}
\usepackage{cleveref}

\usepackage{subcaption}

\begin{document}

\title{eVTOL Aircraft Energy Overhead Estimation under Conflict Resolution in High-Density Airspaces}

\author{\IEEEauthorblockN{Alex Zongo and Peng Wei}
\IEEEauthorblockA{Department of Mechanical and Aerospace Engineering\\
George Washington University\\
Washington, DC, 20052, USA\\
\textit{\{a.zongo, pwei\}@gwu.edu}}
}

\maketitle

\begin{abstract}

Electric vertical takeoff and landing (eVTOL) aircraft operating in high-density urban airspace must maintain safe separation through tactical conflict resolution, yet the energy cost of such maneuvers has not been systematically quantified. This paper investigates how conflict-resolution maneuvers under the Modified Voltage Potential (MVP) algorithm affect eVTOL energy consumption. Using a physics-based power model integrated within a traffic simulation, we analyze approximately \(71,767\)  en route sections within a sector, across traffic densities of \(10-60\) simultaneous aircraft.
The main finding is that MVP-based deconfliction is energy-efficient: median energy overhead remains below $1.5\%$ across all density levels, and the majority of en route flights within the sector incur negligible penalty. However, the distribution exhibits pronounced right-skewness, with tail cases reaching \(44\%\) overhead at the highest densities due to sustained multi-aircraft conflicts. The $95$th percentile ranges from \(3.84\%\) to \(5.3\%\), suggesting that a \(4-5\%\) reserve margin accommodates the vast majority of tactical deconfliction scenarios.
To support operational planning, we develop a machine learning model that estimates energy overhead at mission initiation. Because conflict outcomes depend on future traffic interactions that cannot be known in advance, the model provides both point estimates and uncertainty bounds. These bounds are conservative; actual outcomes fall within the predicted range more often than the stated confidence level, making them suitable for safety-critical reserve planning.
Together, these results validate MVP's suitability for energy-constrained eVTOL operations and provide quantitative guidance for reserve energy determination in Advanced Air Mobility.

\end{abstract}
\begingroup
\renewcommand\thefootnote{}\footnote{
© 2026 IEEE. Personal use of this material is permitted. Permission from IEEE must be obtained for all other uses, in any current or future media, including reprinting/republishing this material for advertising or promotional purposes, creating new collective works, for resale or redistribution to servers or lists, or reuse of any copyrighted component of this work in other works.
}
\addtocounter{footnote}{-1}
\endgroup

\IEEEpeerreviewmaketitle

\section{Introduction}
\subsection{Motivation}

The vision of Advanced Air Mobility (AAM) promises to reshape urban and regional transportation through on-demand aviation services enabled by electric vertical takeoff and landing (eVTOL) aircraft~\cite{faa2023uam}. Characterized by distributed electric propulsion, reduced acoustic signatures, and vertical flight capabilities, these vehicles are positioned to operate from vertiports embedded within metropolitan areas~\cite{nasa2021aam}. Yet as this vision approaches operational reality, a fundamental challenge emerges: the successful integration of large-scale eVTOL fleets into shared airspace demands not only robust separation assurance but also a quantitative understanding of the energy costs that tactical deconfliction entails.

Unlike conventional aircraft, eVTOL platforms operate under stringent energy constraints imposed by current battery technology. The energy density limitations of lithium-ion cells leave little margin for inefficiency~\cite{sripad_venkatasubramanian_energy_efficient_battery}, making precise energy planning essential for mission viability~\cite{taye2024feasibility}. In high-density airspace where dozens of aircraft simultaneously navigate shared corridors, a natural question arises: to what extent do conflict-resolution maneuvers impact energy consumption?

Understanding this relationship is essential, not only to ensure adequate reserves in worst-case scenarios but also to avoid overly conservative planning that unnecessarily constrains capacity and increases operational costs. Traditional flight planning computes energy requirements based on nominal trajectories, leaving operators without quantitative guidance on the additional expenditure that tactical deconfliction may entail. The absence of quantitative characterization of deconfliction energy costs forces a choice between potentially insufficient reserves and excessive conservatism that degrades system efficiency.

Among deterministic approaches for airborne separation assurance, the Modified Voltage Potential (MVP) method has emerged as a widely studied baseline. Originally developed by Eby~\cite{eby1995self} and refined by Hoekstra et al.~\cite{HOEKSTRA_mvp}, MVP iteratively computes minimal velocity adjustments to prevent protected-zone incursions within a specified look-ahead time~\cite{hoekstra_state_based_cd_cr}. Comparative studies show that MVP achieves superior performance in safety and efficiency metrics at high traffic densities~\cite{hoekstra_state_based_cd_cr,ribeiro_joost_hoekstra_review}, and its geometric formulation avoids the explainability challenges of learning-based methods in safety-critical applications. However, while MVP ensures separation, the energy implications of its commanded maneuvers have not been systematically characterized.

\subsection{Related Work}
Research relevant to this study spans three areas: eVTOL energy consumption modeling, tactical conflict resolution, and predictive methods for air traffic management. We review each in turn, highlighting the gap at their intersection.

\textbf{eVTOL Energy Consumption Modeling.} 
Energy consumption modeling for eVTOL aircraft has followed two principal approaches. Physics-based models derive power requirements from rotor aerodynamics and momentum theory, capturing the distinct energy signatures of hover, transition, and cruise flight phases~\cite{Johnson2013,Leishman2006}. Sripad and Viswanathan applied such models to compare eVTOL aircraft configurations, demonstrating that vectored-thrust designs achieve superior cruise efficiency~\cite{sripad_venkatasubramanian_energy_efficient_battery}. Building on physics-based foundations, Taye et al.\ developed mission-level energy models incorporating wind effects and battery state-of-charge dynamics for feasibility assessment and trajectory planning~\cite{taye2025trajectory,taye2024feasibility,thompson2023octocopter}.
At the fleet level, energy-aware traffic management has received growing attention. Taye et al.\ proposed strategic scheduling that accounts for aggregate energy demand~\cite{taye2025strategic}, while Ayalew et al.\ incorporated collision avoidance into reserve planning using conditional value-at-risk (CVaR)~\cite{ayalew2023data}. More recently, Gonzalez and Huynh developed a parametric tilt-rotor eVTOL model in OpenVSP and integrated its performance characteristics into traffic flow management analysis for AAM \cite{victoria_aam_ucirvine}, providing the aircraft configuration data adopted in the present study. However, these works either focus on single-aircraft optimization or treat conflict resolution as a binary event, without characterizing how the intensity and duration of tactical maneuvering continuously affects energy consumption.

\textbf{Tactical Conflict Resolution.}
Separation assurance methods ensure safe spacing when strategic deconfliction is insufficient. The Modified Voltage Potential (MVP) algorithm, introduced by Eby~\cite{eby1995self} and refined by Hoekstra et al.~\cite{HOEKSTRA_mvp}, computes minimal velocity adjustments through geometric analysis of predicted conflicts. Comparative evaluations show MVP achieves low collision rates and minimal path deviation at high traffic densities~\cite{hoekstra_state_based_cd_cr,ribeiro_joost_hoekstra_review}. Alternative approaches include mixed-integer optimization for scheduled deconfliction~\cite{Pelegrn2023UrbanAM} and reinforcement learning~\cite{brittain2019autonomousairtrafficcontroller}, later integrated with demand-capacity balancing~\cite{chen_dcb_deconfliction}. These studies evaluate algorithms primarily on safety metrics such as loss of separation events and near mid-air collision rates, rather than energy consequences. The downstream effect of conflict-resolution maneuvers on energy consumption remains unquantified.

\textbf{Relevant Machine Learning Models.} 
Data-driven methods have been applied to both separation assurance and energy prediction, though rarely together. For separation assurance, deep reinforcement learning has shown promise in high-density scenarios~\cite{brittain_autono_separation_assurance,aziz2026transformerbasedmultiagentreinforcementlearning}. For energy prediction, ensemble methods have achieved accurate power estimates for quadrotor operations~\cite{DAI_modeling_power_consumption_est_quad}. However, these approaches produce point estimates without quantifying uncertainty. Prediction with calibrated confidence bounds, rather than single values, remains underexplored for eVTOL energy consumption, particularly under the stochastic conditions introduced by traffic interactions.

In summary, eVTOL energy models, tactical conflict resolution algorithms, and data-driven prediction methods have each advanced independently, but no prior work has quantified how conflict-resolution maneuvers affect eVTOL energy consumption or provided uncertainty-aware predictions. This paper addresses that gap through the following contributions.

\subsection{Contributions}
This paper addresses the intersection of these three areas by characterizing how MVP-based tactical deconfliction affects eVTOL energy consumption. The contributions are as follows.

\begin{enumerate}
    \item We develop a physics-based eVTOL power consumption model and integrate it with MVP-based tactical deconfliction within a traffic-level simulation, enabling systematic evaluation of energy efficiency across high-density scenarios.
    \item We empirically demonstrate that MVP-based deconfliction is energy-efficient, with median overhead below \(1.5\%\) across all density levels studied, while identifying traffic conditions that produce heavy-tailed overhead distributions relevant to reserve planning.
    \item We develop a machine learning model that quantifies uncertainty in energy overhead during en route flight within a sector. Because conflict outcomes depend on future traffic interactions, precise point prediction is not reliable; the model instead provides calibrated uncertainty bounds that support conservative yet targeted reserve energy determination. 
\end{enumerate}

The remainder of this paper is organized as follows. Section~\ref{sec:evtol_power_consumption_model} presents the eVTOL aircraft power consumption model. Section~\ref{sec:conflict_resolution} describes the conflict resolution and simulation framework. Section~\ref{sec:energy_prediction_model} develops the energy prediction machine learning model. Section~\ref{sec:experiments} evaluates the framework experimentally. Section~\ref{sec:discussion} discusses the results and limitations, and Section~\ref{sec:conclusion} concludes with principal findings and directions for future work.

\section{eVTOL Power Consumption Model}
\label{sec:evtol_power_consumption_model}

The energy consumption model captures tilt-rotor eVTOL aircraft power requirements in cruise flight. The model follows the component build-up methodology established in classical aircraft design \cite{Raymer2018} while incorporating rotorcraft-specific formulations from momentum theory \cite{Johnson2013,Leishman2006}. The baseline configuration represents a six-rotor tilt-rotor eVTOL aircraft in the Joby S4 class, selected for its prominence in the certification pipeline. The geometric and aerodynamic parameters are drawn from the OpenVSP-based parametric model developed by Gonzalez and Huynh \cite{victoria_aam_ucirvine}, who designed and analyzed this tilt-rotor configuration for traffic flow management studies,  supplemented by published data from \cite{Silva2018}. Table~\ref{tab:evtolspec} summarizes key configuration parameters.

Since the traffic simulations in this study operate exclusively in the cruise phase, we present cruise-specific formulations. The simulation framework implements a full-envelope model capable of representing hover and transition phases, but these phases are not the focus of this paper; extension to climb, descent, and holding patterns remains an area for future work.

\subsection{Aerodynamic Drag}
\label{sub:aero_drag}

In cruise flight, the nacelles are oriented horizontally, and the wing fully supports the aircraft's weight. The drag model computes contributions from each major aircraft component: wing, fuselage, tail surfaces, propulsion pods, and landing gear.

The zero-lift drag coefficient for streamlined components follows the standard form:
\begin{align*}
    C_{D,0}^{(i)} = \mathrm{FF}^{(i)} \cdot C_f^{(i)} \cdot \frac{S_{\mathrm{wet}}^{(i)}}{S_{\mathrm{ref}}},
\end{align*}
where $\mathrm{FF}^{(i)}$ is the form factor accounting for thickness effects \cite{Raymer2018,shevell_fundamentals_of_flight}, $C^{(i)}_f=0.455/ \big( \log^{2.58}_{10}(\mathrm{Re}) \big)$ is the skin friction coefficient from the Prandtl-Schlichting turbulent flat-plate formula \cite{shevell_fundamentals_of_flight}, $S_{\mathrm{wet}}^{(i)}$ is the wetted area of component $i$, and $S_{\mathrm{ref}}=S_\mathrm{wing}$, the wing area whose value is given in Table~\ref{tab:evtolspec}~\cite{Raymer2018}. \(\mathrm{Re}\) is the Reynolds number estimated at the set cruising conditions. Landing gear and propulsion pods are modeled using bluff-body correlations~\cite{Hoerner1965}. Given the same reference area used in the individual component drag computation, the total parasitic drag coefficient is~\cite{Raymer2018}:
\begin{align}
    \label{eq:parasite_drag_cruise_coef}
    C_{D,p} = \sum_i C_{D,0}^{(i)}.
\end{align}

The wing also produces induced drag from lift generation. From lifting-line theory \cite{Anderson2017}, the induced drag is:
\begin{align*}
    D_i \approx \frac{W^2}{ \pi \cdot \mathrm{AR} \cdot e \cdot q \cdot S_{\mathrm{wing}}},
\end{align*}
where $\mathrm{AR}$ is the wing aspect ratio, $e=0.8$ is the Oswald efficiency factor. $q=\frac{1}{2}\rho V^2$ is the dynamic pressure where \(\rho\) is the air density at cruise altitude $h=2000$~ft, and \(V\) the airspeed.  
$W$ is the aircraft weight (equal to wing lift in steady level cruise). 
The total drag is $D_{\mathrm{total}} = D_p + D_i$, where $D_p = C_{D,p} \cdot q \cdot S_{\mathrm{ref}}$ with \(C_{D,p}\) computed from \Cref{eq:parasite_drag_cruise_coef} and  $S_\mathrm{ref}=S_\mathrm{wing}$~\cite{anderson1999performance}. Note that the parasite drag is dominant ($D_p \gg D_i$).

\subsection{Rotor Power}
\label{sub:rotor_power}

In cruise, the wing and tail generate enough lift to support the aircraft's weight while the rotors provide forward thrust to overcome drag.

\paragraph{Thrust Requirement}
The thrust required to maintain steady flight at speed $V$ is defined as follows~\cite{anderson1999performance,Raymer2018}:
\begin{align*}
    T_{\mathrm{req}} = D_{\mathrm{total}}.
\end{align*}

\paragraph{Induced Power}
At cruise speeds, the Glauert high-speed approximation applies \cite{Leishman2006}. The induced velocity through the rotor disk is:
\begin{align*}
    v_i \approx \frac{v_h^2}{V},
\end{align*}
where $v_h=\sqrt{T_{\mathrm{req}}/(2\rho A_{\mathrm{total}})}$ is the hover induced velocity and $A_\mathrm{total}=A \cdot N_\mathrm{rotors}$ is the total rotor disk area. The total induced power, accounting for non-ideal losses, is:
\begin{align}
    \label{eq:induced_pw}
    P_{\mathrm{ind}} = \kappa \cdot T_{\mathrm{req}} \cdot v_i,
\end{align}
where $\kappa = 1.15$ captures non-uniform inflow, tip losses, and other rotor non-idealities in cruise \cite{Johnson2013,Leishman2006}.

\paragraph{Profile Power}
The profile power overcomes blade drag and is computed via blade element theory \cite{Johnson2013}:
\begin{align}
    \label{eq:profile_pw}
    P_{\mathrm{profile}} = \frac{C^b_d \sigma}{8} (1 + k_\mu \mu_r^2) \cdot \rho A (\Omega R)^3 \cdot N_{\mathrm{rotors}},
\end{align}
where $\sigma=N_b c /(\pi R)\approx 0.083$ is the rotor solidity, a value within the typical range of  ($0.05-0.12$) for rotorcraft~\cite{Johnson2013}; \(R\) is the rotor radius, $N_b=5$ is the rotor number of blades, $A$ is the single-rotor disk area, and $c$ is the blade chord. 
$\mu_r=V / (\Omega R)$ is the advance ratio where \(\Omega\) is the angular velocity of rotor at cruise. In this study \(\Omega = 300\)~RPM at the set cruise altitude, according to flight simulations of the Joby S4 aircraft in Microsoft Flight Simulator~\cite{joby_msfs_pressrelease_116}.  
$k_\mu=4.65$ accounts for advancing blade effects~\cite{Johnson2013}. The blade drag coefficient $C_{d}^b = C^b_{d,0} \cdot (1 + k_{\mathrm{lift}} \cdot (C_T/\sigma)^2)$ includes lift-dependent contributions, with $C^b_{d,0}=0.012$, $k_{\mathrm{lift}}=6$, and $C_T=T_\mathrm{req}/(\rho \cdot A \cdot (\Omega R)^2 \cdot N_\mathrm{rotors})$ is the thrust coefficient, characterizing the thrust loading per rotor disk~\cite{Johnson2013,Leishman2006}.

\paragraph{Parasite Power}
The power required to overcome airframe drag in forward flight is:
\begin{align}
    \label{eq:parasite_pw}
    P_{\mathrm{parasite}} = D_{\mathrm{total}} \cdot V.
\end{align}

\subsection{Total Power and Energy Integration}

Combining the power components from Equations~\eqref{eq:induced_pw}--\eqref{eq:parasite_pw} with drivetrain losses and auxiliary loads yields:
\begin{align}
    \label{eq:total_power}
    P_{\mathrm{total}} = \frac{P_{\mathrm{ind}} + P_{\mathrm{profile}} + P_{\mathrm{parasite}}}{\eta_{\mathrm{drv}}} + P_{\mathrm{hotel}},
\end{align}
where $\eta_{\mathrm{drv}}=0.85$ represents drivetrain efficiency (motors, inverters, and mechanical transmission)~\cite{Johnson2013} and $P_{\mathrm{hotel}}\approx2$~kW accounts for avionics, environmental control, and auxiliary systems. The energy consumed over a flight segment is:
\begin{align}
    \label{eq:energy_integrated_over_ow}
    E = \int_0^{t_f} P_{\mathrm{total}}(t) \, dt.
\end{align}

\begin{figure}[t]
    \centering
    \includegraphics[width=\linewidth]{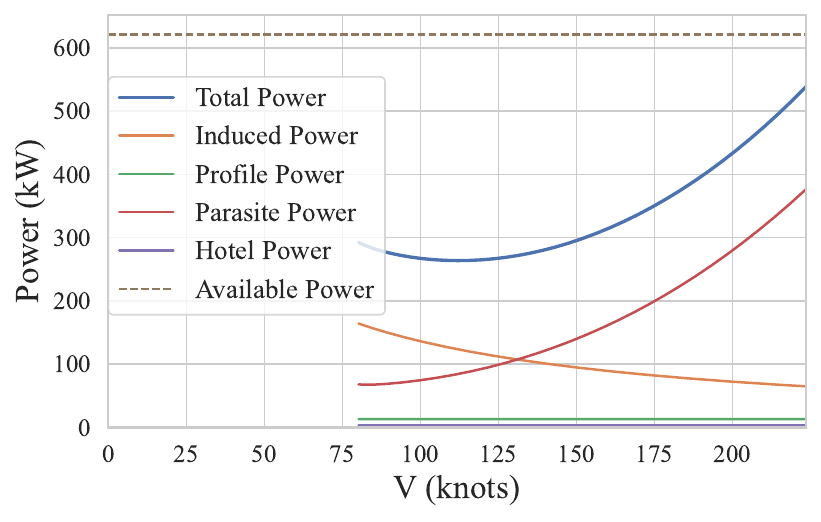}
    \caption{Power-velocity relationship for the baseline tilt-rotor configuration in cruise flight ($V \geq 85$~kt), showing contributions from induced, profile, parasite, and hotel power. The available shaft power (dashed) provides margin throughout the operating envelope.
    }
    \label{fig:power_velocity}
\end{figure}

Fig.~\ref{fig:power_velocity} illustrates the power-velocity relationship for the baseline configuration of our selected eVTOL aircraft Joby S4. The total available shaft power of the $690$~kW (dashed line in Fig.~\ref{fig:power_velocity}) provides a substantial margin throughout the cruise envelope. The curve exhibits qualitative behavior consistent with published tilt-rotor analyses \cite{Silva2018,victoria_aam_ucirvine}.

\subsection{Model Scope and Limitations}

Several aspects of the model warrant discussion. First, the configuration parameters in Table~\ref{tab:evtolspec} are primarily derived from the OpenVSP parametric model from \cite{victoria_aam_ucirvine} supplemented by published data \cite{Silva2018,joby_msfs_pressrelease_116}, and typical rotorcraft design practices \cite{Johnson2013} where specific values were unavailable. Second, the model has not been validated quantitatively against flight test data; validation is limited to qualitative agreement with expected power-velocity trends for tilt-rotor aircraft~\cite{victoria_aam_ucirvine}. Third, this study focuses on cruise operations; extensions to full-envelope operations---including climb, descent, and holding patterns---remain an area for future work requiring additional validation data.

\begin{table}[t]
\centering
\caption{Baseline eVTOL Configuration Parameters. See \cite{victoria_aam_ucirvine,Silva2018,Johnson2013} for additional details.}
\label{tab:evtolspec}
\resizebox{\columnwidth}{!}{%
\begin{tabular}{l c l c}
\toprule
\textbf{Parameter} & \textbf{Value} & \textbf{Parameter} & \textbf{Value} \\
\midrule
Number of rotors     & 6      & Wing aspect ratio      & 10.8 \\
Rotor diameter (m)   & 2.9    & Max takeoff mass (kg)  & 2{,}177 \\
Wing area (m$^2$)    & 10.83  & Battery capacity (kWh) & 136  \\
Wing airfoil geometry & NACA2418 & Wing span (m) & 10.66 \\
Max RPM      & 800  & \# of blades per rotor  & 5 \\
Cruise RPM   & 300  & Cruise altitude (ft) & 2,000 \\
Max shaft power (kW) & $115 \times 6$ & Peak motor torque (Nm) & 1{,}677.15 \\
\bottomrule
\end{tabular}
}
\end{table}

\section{Conflict Resolution and Simulation Framework}
\label{sec:conflict_resolution}
This section describes the simulation environment used to generate training data for the predictive energy model. We present the airspace scenario, the MVP tactical deconfliction scheme, and the data collection methodology. 

\subsection{Airspace and Traffic Scenario}
\label{subsec:airspace_traffic_scene}
\begin{figure}
    \centering
    \includegraphics[width=\linewidth]{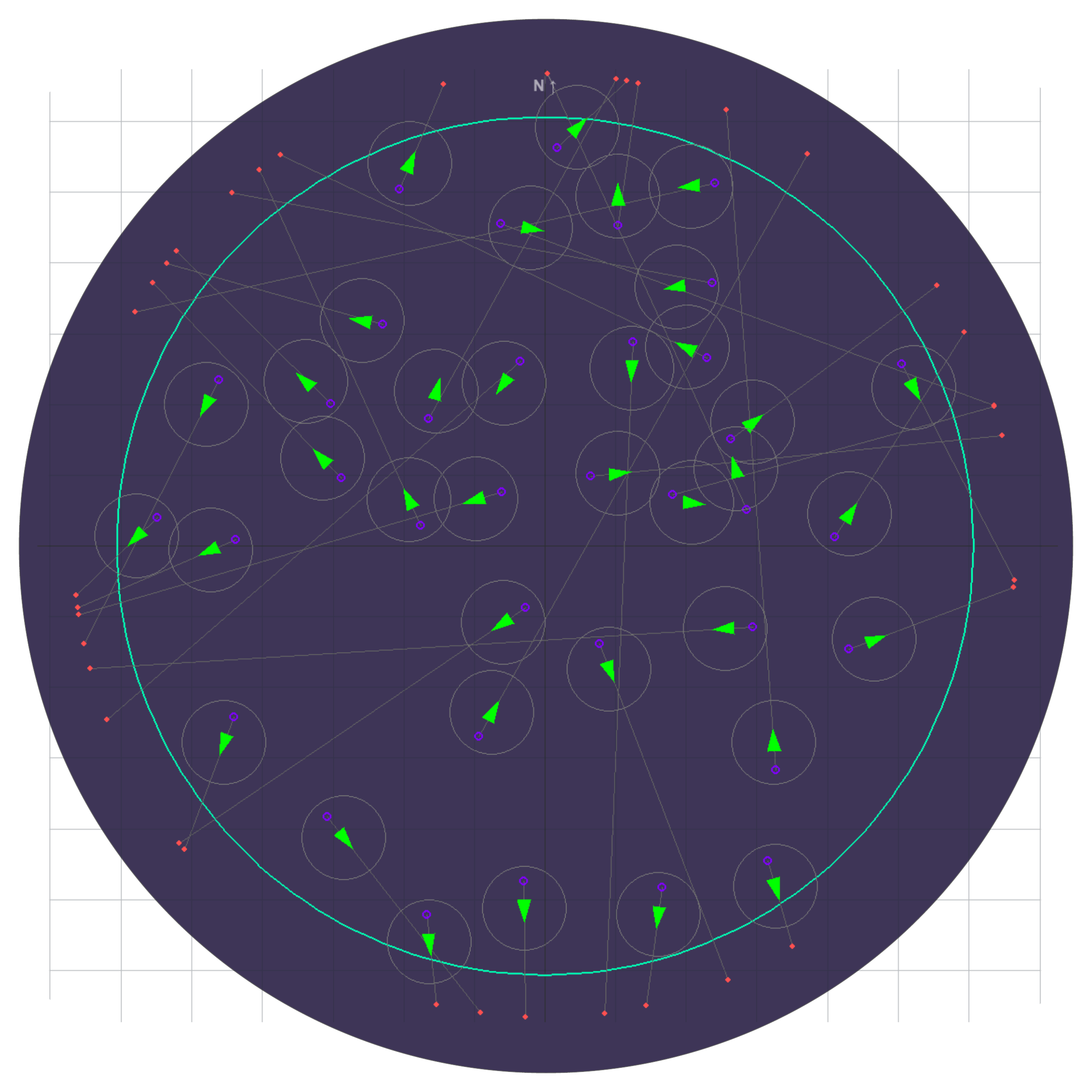}
    \caption{
    A traffic-level scenario where the eVTOL aircraft (in green) are cruising, inside a \(10\)~nm radius airspace sector, towards their exit (red dots). The circle around each aircraft represents the protected zone. (see supplementary video \protect\footnotemark)
    }
    \label{fig:traffic_scenario}
\end{figure}

The simulation environment consists of a circular free-flight airspace sector where each eVTOL aircraft operates without a predefined route, as shown in Fig.~\ref{fig:traffic_scenario}. This configuration, common in separation assurance research \cite{freeflight_hoekstra}, provides a control setting in which traffic density and conflict frequency can be systematically varied while isolating the effects of tactical deconfliction from strategic flow management. 

\subsubsection{Sector Geometry} 
The airspace is defined as a circular sector of radius $R_{\text{sector}}=10$ nm centered at a fixed coordinate origin. This dimension is representative of inter-vertiport spacing in metropolitan AAM networks; studies of vertiport placement for cities such as New York employ node separations of approximately $16$ km to serve primary demand areas \cite{xuxi_uam_nyc}, yielding transit distances consistent with a circular sector at this scale.
All operations occur at a single cruise altitude $h=2000$~ft, eliminating vertical separation as a resolution dimension and focusing the analysis on horizontal conflict resolution.

\subsubsection{Traffic Generation} 
At the start of each simulation run, $N$ aircraft are instantiated simultaneously within the sector. Initial positions are sampled uniformly around the sector circumference and randomly scaled radially inward so that all aircraft begin inside the boundary. To prevent immediate loss-of-separation events at initialization, a minimum clearance constraint is enforced: each aircraft must be separated from all others by a distance proportional to the sector radius and inversely related to traffic density. Specifically, the minimum initial separation $d_{\text{min}}$ is computed as:
\begin{align*}
    d_{\text{min}} = \alpha \cdot \frac{R_{\text{sector}}}{\sqrt{N}}
\end{align*}
where $\alpha=0.9\sqrt{\pi}$ is a scaling factor ensuring adequate initial spacing. All aircraft are initialized at the maximum range speed $V_{\text{br}}$ defined as the speed minimizing energy consumption per unit distance. 
\begin{align*}
    V_{\text{br}} = \arg \min \frac{P(V)}{V}
\end{align*}
From the baseline eVTOL aircraft configuration, the power model described in Section~\ref{sec:evtol_power_consumption_model} and the representative power-velocity curve in Fig.~\ref{fig:power_velocity}, $V_{\text{br}}\approx157$~kt.

\subsubsection{Destination Assignment}
Each aircraft is assigned an exit waypoint located on the sector boundary. The exit bearing is sampled uniformly from the range \([60^\circ, 180^\circ]\) relative to the aircraft's entry bearing, with the direction (clockwise or counterclockwise) chosen randomly with equal probability. This angular range produces a distribution of conflict geometries---including crossing angles from shallow to perpendicular---representative of the encounter types expected in unstructured urban airspace. 

\subsubsection{Density Parameterization}
Traffic density is controlled directly through the number of simultaneously cruising aircraft $N$. Traffic counts range from $N=10$ to $N=60$, spanning instantaneous densities from approximately $32$ to $191$ aircraft per $1000 \ \mathrm{nm}^2$. The upper bound corresponds to near-saturation conditions where the initial spacing constraint approaches the protected zone radius, ensuring that the aircraft initial positions remain geometrically feasible while generating challenging conflict scenarios representative of projected high-density AAM operations. This batch-spawn approach ensures that all aircraft experience the same initial traffic environment, enabling controlled comparison of energy outcomes across density levels.

\subsubsection{Conflict Potential}
The combination of inward-scaled initial positions, randomized entry bearings, and constrained exit angles ensures that aircraft trajectories converge toward the sector interior, creating a natural conflict zone. The geometry maximizes pairwise conflict frequency while maintaining realistic encounter angles, enabling efficient data collection across a range of conflict scenarios. 

\begin{table}[t]
\centering
\caption{Simulation Environment Parameters}
\label{tab:sim_env_params}
\begin{tabular}{l c l}
\toprule
\textbf{Parameter} & \textbf{Symbol} & \textbf{Value} \\
\midrule
Sector radius              & $R_{\mathrm{sector}}$ & 10 nm \\
Max-range cruise speed    & $V_{\mathrm{br}}$     & 157 kt \\
Initial separation factor  & $\alpha$              & $0.9\sqrt{\pi} \approx 1.60$ \\
Exit bearing range         & ---                   & $[60^\circ, 180^\circ]$ \\
Traffic count              & $N$                    & $10-60$ \\
\bottomrule
\end{tabular}
\end{table}

\footnotetext{\href{https://youtu.be/L_CJrEOIxFg}{Supplementary video (https://youtu.be/L\_CJrEOIxFg)}}

\subsection{MVP Tactical Deconfliction}
\label{subsec:mvp_resolution}
The Modified Voltage Potential (MVP) algorithm, introduced in Section~I, provides decentralized, deterministic conflict resolution based on geometric analysis of predicted conflicts~\cite{freeflight_hoekstra,eby1995self,HOEKSTRA_mvp}. This section details the implementation used in our simulations. 

\subsubsection{Conflict Detection}
A conflict is detected when two aircraft are predicted to lose separation within a specified look-ahead time $t_{\mathrm{look}}$. Each aircraft maintains a protected zone, i.e., a horizontal disc of radius $r_\mathrm{pz}$ centered on its position. The conflict detection condition is:
\begin{align*}
    ||\mathbf{d}_{\mathrm{CPA}} || < r_\mathrm{pz}, \quad \mathrm{and} \quad t_{\mathrm{CPA}} < t_{\mathrm{look}}
\end{align*}
where \(\mathbf{d}_{\mathrm{CPA}}\) is the predicted relative position at the closest point of approach (CPA) and \(t_{\mathrm{CPA}} \) is the time to CPA. 

The look-ahead time is set to \(t_{\mathrm{look}}=90 \ \mathrm{s}\), providing sufficient margin for detection across all encounter geometries while remaining within the tactical deconfliction timeframe of $60-180$~s typical of airborne self-separation systems~\cite{nasa2020tactical,nasa2025tactical}.

\subsubsection{Resolution Vector Computation} 
Upon conflict detection, MVP computes a velocity adjustment that eliminates the predicted protected-zone intrusion. The algorithm determines the minimum displacement required to move the predicted CPA position to the boundary of the intruder's protected zone, then converts this spatial displacement into a velocity change~\cite{ribeiro_joost_hoekstra_review, hoekstra_state_based_cd_cr}:
\begin{align}
\Delta v =\frac{r_{\mathrm{pz}} - ||\mathbf{d}_{\mathrm{cpa}}||}{t_{\mathrm{cpa}}} \cdot \hat{\mathbf{n}}
\end{align}
where \( \hat{\mathbf{n}} \) is the unit vector from the predicted intruder's position at CPA toward the nearest point on the protected-zone boundary. This formulation yields the smallest velocity change that resolves the conflict, resulting in minimal path deviation \cite{hoekstra_state_based_cd_cr,eby1995self,HOEKSTRA_mvp}.

\subsubsection{Multi-Conflict Resolution and Coordination}
Given a pair of conflicting aircraft, both compute avoidance vectors pointing away from each other's predicted CPA positions, producing inherently complementary maneuvers without explicit communication~\cite{hoekstra_state_based_cd_cr}. When an aircraft detects multiple simultaneous conflicts, MVP sums the individual resolution vectors. The algorithm operates iteratively, recomputing resolutions at each time step as the traffic situation evolves~\cite{HOEKSTRA_mvp,hoekstra_state_based_cd_cr,ribeiro_joost_hoekstra_review}.

\subsubsection{Recovery}
After executing a resolution maneuver, the aircraft reverts to its original velocity when two criteria are met~\cite{Schaberg2020}: (1) the original conflict has been resolved, indicated by passage beyond the computed CPA; and (2) resuming the original velocity would not create a new conflict within the look-ahead horizon.

\begin{figure}
    \centering
    \includegraphics[width=\linewidth]{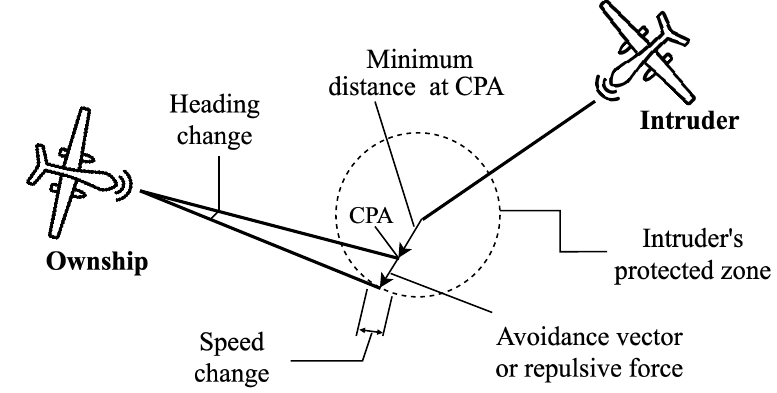}
    \caption{Modified Voltage Potential (MVP) conflict resolution geometry adapted from \cite{HOEKSTRA_mvp}.}
    \label{fig:mvp_representation}
\end{figure}

\begin{table}[t]
\centering
\caption{Conflict Detection and Resolution Parameters}
\label{tab:conf_detect_reso_params}
\begin{tabular}{l c l}
\toprule
\textbf{Parameter} & \textbf{Symbol} & \textbf{Value} \\
\midrule
Protected zone radius              & $r_{\mathrm{pz}}$ & 0.6 nm \\
Permitted speed range & --- & 85-185 kt \\
Look-ahead time    & $t_{\mathrm{look}}$     & \(90\) s \\
\bottomrule
\end{tabular}
\end{table}

\subsection{Data Collection}
\label{subsec:data_collection}
The simulation framework is implemented on top of the BlueSky open-source air traffic simulator \cite{Hoekstra_bluesky}, extended with the eVTOL power consumption model described in Section \ref{sec:evtol_power_consumption_model}. Each simulation run initializes $N$ aircraft according to the traffic scenario defined in Section \ref{subsec:airspace_traffic_scene} and executes MVP-based tactical deconfliction until all aircraft exit the sector. 

\subsubsection{Experimental Design}
Simulations are conducted across \(11\) traffic density levels with $N \in \{ 10,15,20,25,30,35,40,45,50,55,60\}$ aircraft. For each traffic count, $200$ independent runs are executed with randomized initial positions and destination assignments. The resulting dataset comprises about $71,767$ aircraft transits spanning a range of conflict intensities, from sparse traffic with minimal deconfliction activity to congested conditions requiring continuous MVP intervention.

\subsubsection{Energy Metrics}
The primary outcome variable is relative energy overhead due to conflict resolution. For each transit, we compute baseline energy consumption $e_{\mathrm{baseline}}$ along a conflict-free trajectory at $V_{\mathrm{br}}$, and actual consumption $e_{\mathrm{actual}}$ by integrating instantaneous total power over the realized trajectory and dividing by path length. The relative overhead is:
\begin{align}
\label{eq:energy_overhead}
    \delta_E = \frac{e_{\mathrm{actual}} - e_{\mathrm{baseline}}}{e_{\mathrm{baseline}}},
\end{align}
expressing overhead as a fractional increase relative to the conflict-free baseline, enabling comparison across routes of different lengths. 

\subsubsection{Feature Extraction} 
To support the predictive model developed in Section \ref{sec:energy_prediction_model}, we extract features available at mission initiation---before the aircraft begins its transit. Summarized in Table~\ref{tab:input_features}, these features characterize the aircraft's initial state, the surrounding traffic, and the conflict situation as assessed by the MVP algorithm at the beginning of cruise. All features are normalized to facilitate model training across varying traffic densities. 

\begin{table}[t]
\centering
\caption{Input Features for Energy Prediction Model}
\label{tab:input_features}
\begin{tabular}{p{2.2cm} p{5.5cm}}
\toprule
\textbf{Category} & \textbf{Features} \\
\midrule
Aircraft State & Normalized speed deviation from $V_{\mathrm{br}}$; radial position from sector center; route distance; the variance of relative bearing and speed among neighbors; angle between ownship's planned heading and mean neighbor heading \\
\addlinespace
MVP Resolution & Suggested heading change $\Delta\psi$; suggested speed change magnitude $||\Delta v||$; sector congestion $N/(\pi R_{\mathrm{sector}}^2)$; neighborhood conflict density \\
\addlinespace
Conflict Severity & Weighted urgency measure $w_{\mathrm{deg}}$ combining temporal and spatial proximity; minimum $t_{\mathrm{CPA}}$; minimum $d_{\mathrm{CPA}}$ \\
\bottomrule
\end{tabular}
\end{table}

Not all detected conflicts pose equal risk to energy consumption. A conflict that is seconds away and deeply intruding into the protected zone demands aggressive maneuvering and thus higher energy cost, whereas a distant conflict may resolve with minimal intervention. To capture this distinction, we define a conflict severity metric $w_{\mathrm{deg}}$ that assigns higher weight to conflicts that are both temporally imminent and spatially severe:

{\small
\begin{align*}
    w_{\mathrm{deg}} = \frac{1}{N-1} \sum_{i\in \mathcal{C}} \exp \Big( - \frac{t_{\mathrm{CPA},i} }{0.35 t_{\mathrm{look}}} \Big) \cdot \max \Big(   \frac{ r_{\mathrm{pz}}-d_{\mathrm{CPA},i} }{r_{\mathrm{pz}}},0  \Big).
\end{align*} }
The exponential term penalizes temporal proximity, while the linear term scales with the depth of predicted protected-zone intrusion. The normalization by $N-1$ enables comparison across different traffic densities.

\section{Energy Overhead Prediction}
\label{sec:energy_prediction_model}
This section presents a machine learning model that estimates the distribution of energy overhead at mission initiation, providing uncertainty bounds for reserve determination.

\subsection{Problem Formulation}
\label{subsec:pred_problem_formulation}
The prediction task is to estimate the relative energy overhead \( \delta_E \) defined in Section \ref{subsec:data_collection}, conditioned on the feature vector $\mathbf{x}$ available at the start of cruise. Since conflict resolution outcomes are inherently stochastic---depending on the evolving behavior of surrounding traffic---we seek a model that captures the distribution \( p(\delta_E\,|\,\mathbf{x}) \) rather than a single point estimate. 

In practice, energy overhead \(\delta_E\) is non-negative and right-skewed: most transits incur small overhead while a few experience substantially higher values due to cascading conflicts in dense traffic. Standard regression models assume symmetric, unbounded targets and are poorly suited to this structure. To address this, we apply a log-transformation: \(z = \log(1 +\delta_E)\). This maps the non-negative, skewed overhead values onto an unbounded scale where a Gaussian assumption is appropriate: \(z \, | \, \mathbf{x} \sim \mathcal{N}( \mu_z(\mathbf{x}), \sigma_z^2(\mathbf{x}) )\).
The model predicts the mean and variance of $z$; energy overhead is then recovered by inverting the transformation: \(\delta_E=\exp(z) - 1\). This approach lets the model output not just a point estimate but also upper and lower bounds that reflect how confident the prediction is for a given set of initial conditions.

\subsection{Network Architecture}
\label{subsec:network_architecture}
\begin{figure}[t]
    \centering
    \includegraphics[width=\linewidth]{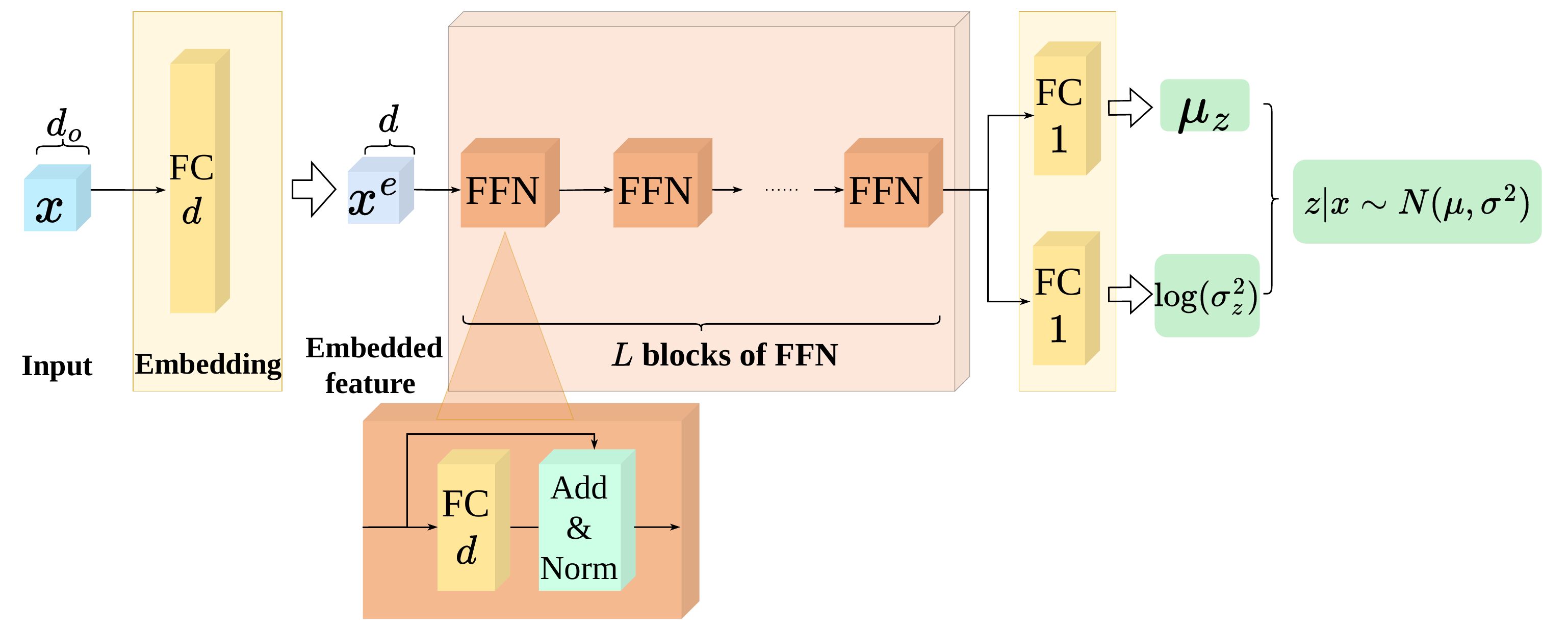}
    \caption{Architecture of the energy prediction neural network. Fully connected (FC) layers with feed-forward network (FFN) blocks and residual connections produce distributional parameters for the transformed energy overhead variable.}
    \label{fig:nn_architecture}
\end{figure} 

The model architecture, illustrated in Fig.~\ref{fig:nn_architecture}, consists of an input embedding layer, a stack of $L$ feedforward (FFN) blocks with residual connections and layer normalization, and two output heads.
The input feature vector $\mathbf{x}\in \mathbb{R}^{d_0}$ is projected into a $\mathrm{d}$-dimensional hidden representation through a linear layer with SiLU activation and dropout. This representation passes through $L$ identical FFN blocks, each applying a residual connection with post-normalization to enable deep feature transformation while maintaining stable gradients. The final hidden state $\mathbf{h}_L$ is mapped to distributional parameters through two linear output heads:
\begin{align*}
    \mu_z &= \mathbf{w}_\mu^\top \mathbf{h}_L + \mathbf{b}_\mu, \\
    \log \sigma_z^2 &= \mathrm{clamp}( \mathbf{w}_\sigma^\top \mathbf{h}_L + \mathbf{b}_\sigma, \lambda_{\mathrm{min}}, \lambda_{\mathrm{max}} ).
\end{align*}
The log-variance is clamped to prevent numerical instability during training. Table~\ref{tab:model_config_hyperparams} summarizes the architecture configuration and training hyperparameters.

\subsection{Training Objective}
\label{subsec:training_objective}
The model is trained by minimizing the Gaussian negative log-likelihood in the transformed space over the neural network parameters $\mathbf{\theta}$:

\begin{equation}
    \label{eq:training_objective}
    \mathcal{L}(\mathbf{\theta}) = \frac{1}{2} \mathbb{E}_{(\mathbf{x},\delta_E)\sim \mathcal{D}} \Bigg[  \log\sigma_z^2(\mathbf{x}) +\frac{(z - \mu_z(\mathbf{x}))^2}{\sigma_z^2(\mathbf{x})} \Bigg],
\end{equation}
where \( z = \log(1+\delta_E) \) is computed from the observed energy overhead. Training in the transformed space improves numerical stability compared to directly modeling the bounded variable $\delta_E$. 
The dataset is partitioned into training (\( 80\% \)) and validation ( \(20\%\)) sets. Optimization uses AdamW with learning rate \(3\times10^{-4}\) and weight decay \(10^{-4}\). Gradient norms are clipped to \(1.0\) to stabilize training. The model checkpoint with the lowest validation NLL is retained for evaluation.

\subsection{Inference}
At inference time, the model outputs the predicted distribution parameters \(\mu_z, \sigma_z^2\) for a given feature vector \(\mathbf{x}\). From these, we derive point estimates and prediction intervals.

\paragraph{Mean Prediction}
The expected energy overhead under the shifted log-normal distribution:
\begin{align}
\label{eq:overhead_pred_estimate}
    \mathbb{E}[\delta_E] \approx \exp\Big( \mu_z + \frac{\sigma_z^2}{2}  \Big) - 1.
\end{align}

\paragraph{Quantile Predictions}
For reserve determination, we compute quantiles by transforming the corresponding Gaussian quantiles:
\begin{align}
    \label{eq:overhead_quant_est}
    \delta_{E,q} \approx \exp\Big(\mu_z + \sigma_z \cdot \Phi^{-1}(q) \Big) - 1,
\end{align}
where $\Phi^{-1}(q)$ is the standard normal quantile function. The upper quantile (e.g., \( q=0.9 \)) provides a conservative estimate for energy reserve calculations, while the prediction interval width quantifies uncertainty. 

\subsection{Evaluation Metrics}
\label{subsec:evaluation_metric}

The model performance is assessed through metrics addressing point prediction accuracy, distributional calibration, and residual diagnostics.

\paragraph{Point Prediction Accuracy}
Point prediction is evaluated in \(\delta_E-\mathrm{space}\) using the predictive mean. We report the mean absolute error (MAE) and root mean squared error (RMSE).

The coefficient of determination \(R^2\)  quantifies the fraction of variance explained by the model relative to a constant mean baseline; \(R^2=1\) indicates perfect prediction, $R^2 = 0$ indicates that the performance is no better than predicting the sample mean, and $R^2 < 0$ indicates higher squared error than the mean baseline, which is a possible outcome when the target variable exhibits high intrinsic stochasticity that input features cannot explain.

The Pearson correlation coefficient measures linear association between predictions and observations, capturing whether predictions track relative ordering even when magnitudes differ.

\paragraph{Uncertainty Calibration}
For well-calibrated uncertainty bounds, the empirical coverage of a nominal prediction interval should match its stated level (e.g., \(90\%\) of observations fall within a 90\% interval). Overcoverage indicates conservative bounds; undercoverage indicates overconfident predictions. 

Mean prediction interval width quantifies sharpness: narrower intervals are preferable when coverage is maintained, as they provide more precise uncertainty bounds.

\paragraph{Residual Diagnostics}
Standardized residuals in the transformed $z$-space assess distributional fit. For a correctly specified model, these residuals should follow a standard normal distribution with mean zero and unit variance.

\begin{table}[t]
\centering
\caption{Model Configuration and Training Hyperparameters}
\label{tab:model_config_hyperparams}
\begin{tabular}{l l}
\toprule
\textbf{Component} & \textbf{Value} \\
\midrule
Input dimension       & \(d_0\) (number of features) \\
Hidden dimension      & \(128\)  \\
FFN blocks            &  \(4\) \\
Activation            & SiLU \\
Dropout               & \(0.05\) \\
Log-variance clamp    & \([-8, 3]\) \\
Optimizer             & AdamW \\
Learning rate         & \(3\times 10^{-4}\) \\
Weight decay          & \(10^{-4}\) \\
Batch size            & 256 \\
Gradient clipping     & 1.0 \\
Epochs                & 10,000 \\
Train/validation split & \(80\% / 20 \%\) \\
Model selection       & Minimum validation NLL \\
\bottomrule
\end{tabular}
\end{table}

\section{Experimental Evaluation}
\label{sec:experiments}

We evaluate the framework in three stages: first, empirical characterization of energy overhead across density levels; second, point-prediction performance of the machine learning model; and third, the calibration of the model's uncertainty bounds, most critical for operational use. 

\subsection{Energy Overhead Characterization}
\label{subsec:energy_overhead_characterization}
\begin{figure}[t]
    \centering
    \includegraphics[width=\linewidth]{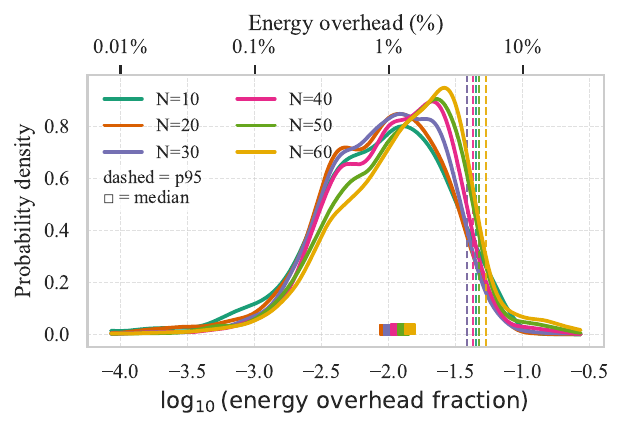}
    \caption{
    Distribution of relative energy overhead by traffic density, shown in log-space to reveal structure obscured by right-skew. Dashed vertical lines indicate \(95\)th Percentile (P95) for each scenario. Medians cluster tightly near $1\%$ across all densities, while P95 values spread from $4.5\%$ to $5.3\%$, illustrating that density primarily affects tail behavior rather than typical energy overhead.
    }
    \label{fig:overhead_dist}
\end{figure}
\begin{figure}[t]
    \centering
    \includegraphics[width=\linewidth]{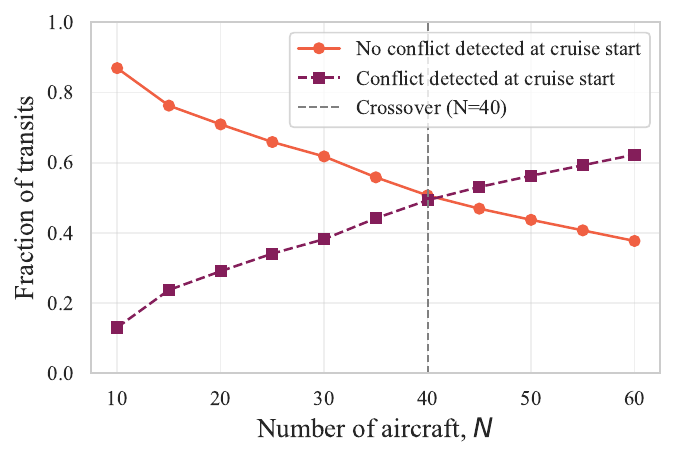}
    \caption{
    Fraction of aircraft detected in conflict by MVP versus conflict-free at the beginning of cruise flight, as a function of traffic density. At low densities, the majority of aircraft begin their en route flight with no active conflict detections. The crossover near $N = 40$ coincides with the onset of heavier energy overhead tails in Fig.~\ref{fig:overhead_dist}.
    }
    \label{fig:conflict_transit_fraction}
\end{figure}

\begin{table}[t]
\centering
\caption{Energy Overhead Statistics by Traffic Density. P90 and P95 represent the 90th and 95th percentiles, respectively.}
\label{tab:energy_overhead_stats}
\begin{tabular}{c c c c c c}
\toprule
\textbf{N} & \textbf{Mean (\%)} & \textbf{Median (\%)} & \textbf{P90 (\%)} & \textbf{P95 (\%)} & \textbf{Max (\%)} \\
\midrule
10 & 1.42 & 0.98 & 3.36 & 4.51 & 6.98 \\
20 & 1.34 & 0.94 & 3.06 & 3.84 & 12.56 \\
30 & 1.43 & 1.00 & 3.07 & 3.87 & 24.87 \\
40 & 1.61 & 1.13 & 3.52 & 4.38 & 22.97 \\
50 & 1.85 & 1.29 & 3.76 & 4.72 & 36.49 \\
60 & 2.12 & 1.44 & 4.06 & 5.30 & 44.23 \\
\bottomrule
\end{tabular}
\end{table}

Fig.~\ref{fig:overhead_dist} presents the distribution of relative energy overhead \(\delta_E \) across traffic density levels, plotted in \(\log\mathrm{-space}\) to reveal structure obscured by the strong right-skew in raw data. Summary statistics are provided in Table~\ref{tab:energy_overhead_stats}. The dataset comprises approximately \(71,767\) sector transits spanning $N\in\{ 10, 15,\dots,60 \}$ aircraft. Notably, no near mid-air collisions (NMACs) occurred across all simulations.

A key finding is that MVP-based deconfliction is energy-efficient: the distributions share a common mode near \(1\%\) overhead across all density levels, and median overhead remains below \(1.5\%\) even at the highest densities. The majority of transits incur a negligible energy penalty from tactical conflict resolution, validating MVP's suitability for energy-constrained eVTOL operations.  

However, the distributions exhibit density-dependent right tails visible in the shifting P95 markers (dashed lines). At \(N = 10\), the \(95\)th percentile falls near \(4.51\%\) while at \(N = 60\) it extends to \(5.3\%\). Maximum observed energy overhead reaches \(44\%\) at \(N=60\) compared to \(7\%\) at \(N=10\). These tail cases arise when aircraft encounter sustained multi-aircraft conflicts requiring prolonged maneuvering. 

The growing tail weight is explained by the increasing fraction of aircraft that are already in detected conflict at the start of cruise. As shown in Fig. \ref{fig:conflict_transit_fraction}, the fraction of aircraft beginning cruise with no MVP-detected conflicts decreases from \(87\%\) at \(N=10\) to \(38\%\) at \(N=60\), with a crossover near \(N=40\). Beyond this threshold, the majority of aircraft begin their en route flight with at least one conflict, increasing the likelihood of compounded interactions that drive the tail energy overhead observed. 

For operational planning, these results support targeted rather than blanket reserve policies. A \(4-5\%\) energy margin accommodates \(95\%\) of tactical deconfliction scenarios across all density levels studied.

\subsection{Prediction Performance}
\label{subsec:prediction_performance}
Having characterized the empirical distribution of energy overhead, we now evaluate whether a machine learning model can provide useful pre-flight estimates of this quantity for individual transits. 

Table~\ref{tab:prediction_metrics} summarizes prediction performance on the held-out validation set. The mean absolute error (MAE) is \(1.0\%\), indicating point predictions are typically within \(1.0\) percentage points of the actual energy overhead. The root mean squared error (RMSE) of \(1.8\%\) reflects larger errors on difficult cases.

The model's primary purpose is to provide calibrated uncertainty bounds for reserve planning, not precise point forecasts. This distinction is important because the eventual energy overhead depends on future traffic encounters and resolution geometry, which are factors that cannot be observed at mission start. Accordingly, we evaluate the model first on its uncertainty calibration (Table~\ref{tab:calibration_metrics}), where it performs well, and report point-prediction metrics for completeness.

The Pearson correlation of 0.50 indicates a moderate linear association between predictions and observations, confirming that the model captures a meaningful signal about which transits will experience higher overhead. The coefficient of determination \(R^2 \approx 0.22\) indicates that the model explains approximately $22\%$ of the variance in energy overhead, which is a reasonable result given that conflict resolution outcomes depend on future traffic interactions that cannot be fully anticipated at pre-flight. While these point-prediction metrics demonstrate useful discriminative capability, the model's primary operational value lies in its calibrated uncertainty bounds, which Table~\ref{tab:calibration_metrics} evaluates.

Stratifying results by whether the aircraft encountered a conflict reveals an expected pattern: MAE is $0.7\%$ for conflict-free transits versus $1.3\%$ for conflict-involved transits. Conflict-free flights follow near-nominal trajectories with little variability, making them easier to predict. Conflict-involved transits exhibit the stochastic maneuvering behavior that limits predictability.

A natural question is whether the model provides value beyond a simple density-conditioned lookup using the empirical statistics in Table~\ref{tab:energy_overhead_stats}. The model's advantage is that it conditions on the full feature vector, including the aircraft's specific initial conflict state, spatial position, and MVP initial resolution parameters, enabling instance-level uncertainty bounds tailored to each transit rather than statistics that treat all aircraft at a given density identically. This distinction is operationally relevant: two aircraft in the same density environment but with different conflict exposures at the start of cruise should receive different reserve recommendations.

\begin{table}[t]
\centering
\caption{Prediction Performance on Validation Set}
\label{tab:prediction_metrics}
\begin{tabular}{l c}
\toprule
\textbf{Metric} & \textbf{Value} \\
\midrule
MAE (overall) & $1.0\%$ \\
MAE (conflict-free) & $0.7\%$ \\
MAE (conflict-involved) & $1.3\%$ \\
RMSE & $1.8\%$ \\
Pearson correlation & $0.50$ \\
$R^2$ & $0.22$ \\
\bottomrule
\end{tabular}
\end{table}

\subsection{Uncertainty Calibration}
While point prediction is limited by inherent unpredictability, the model provides useful uncertainty estimates for reserve planning. Table~\ref{tab:calibration_metrics} reports calibration diagnostics.

The empirical coverage of $96\%$ for the nominal 80\% interval, and $98\%$ for the nominal 90\% interval, indicates the model is conservative---actual overhead falls within predicted bounds more often than the nominal rate suggests. For safety-critical reserve planning, this conservatism is desirable: operators can trust stated bounds will be met at least as often as advertised.

The prediction interval widths of approximately $5.3\%$ ($80\%$ level) and $6.8\%$ ($90\%$ level) are narrow enough to be operationally useful. An operator targeting $90\%$ confidence could use the upper bound of the $90\%$ interval as a reserve margin, obtaining values consistent with the empirical P90 statistics in Table~\ref{tab:energy_overhead_stats}.

The standardized residuals in $z$-space have mean near zero ($0.08$) and standard deviation near one ($0.68$), indicating slight conservatism in the model's internal representation. The model tends to overestimate uncertainty, producing residuals with lower-than-expected variance.

\begin{table}[t]
\centering
\caption{Uncertainty Calibration on Validation Set}
\label{tab:calibration_metrics}
\begin{tabular}{l c c}
\toprule
\textbf{Metric} & \textbf{Nominal} & \textbf{Empirical} \\
\midrule
80\% interval coverage & $80\%$ & $96\%$ \\
90\% interval coverage & $90\%$ & $98\%$ \\
80\% interval width & --- & $5.3\%$ \\
90\% interval width & --- & $6.8\%$ \\
$z$-residual mean & $0$ & $0.08$ \\
$z$-residual std & $1$ & $0.68$ \\
\bottomrule
\end{tabular}
\end{table}

\section{Discussion}
\label{sec:discussion}
\subsection{Interpretation of Results}

The finding that MVP imposes modest typical overhead while maintaining zero NMACs has practical implications for Advanced Air Mobility. Tactical conflict resolution need not be viewed as a significant energy burden, addressing a concern accompanying proposals for high-density urban airspace. The pronounced right-skewness of the overhead distribution, however, warrants attention: while typical transits incur negligible penalty, occasional high-overhead events (reaching $44\%$ at the highest densities) are operationally significant for reserve planning. The $95$th percentile guidance of $4-5\%$ provides a concrete, empirically-grounded reserve target.

The machine learning model provides both useful point predictions ($R^2=0.22$, $\mathrm{Pearson}=0.50$) and well-calibrated uncertainty bounds. For operational planning, the latter remains most critical. This distinction matters for practitioners: the model provides calibrated bounds that support conservative yet targeted reserve determination, rather than forecasting exact values, which the stochastic nature of conflict resolution renders not reliable. The observed overcoverage ($96\%$ empirical at $80\%$ nominal) is acceptable and even desirable for safety-critical applications.

From an operational standpoint, the model enables a shift from blanket to situation-specific reserve policies. Rather than applying a uniform \(5\%\) energy reserve to all flights regardless of conditions, a dispatch system could query the model with pre-flight features such as current traffic count, initial conflict detections, and planned route geometry, and receive tailored upper bounds. For example, an aircraft launching into a low-density sector with no initially detected conflicts might receive a \(90\%\) upper bound near \(3\%\), while one entering a congested sector with multiple initial conflicts might receive a bound near \(7\%\). 
The observed overcoverage means these bounds are conservative in practice: an operator using the $90\%$ prediction interval as a reserve target would find the actual overhead falls within that bound at least $98\%$ of the time.

\subsection{Limitations}

Several aspects warrant discussion. First, the eVTOL power model has not been validated against flight test data; absolute energy values should be interpreted with caution. However, because overhead is computed as a ratio between conflict-affected and conflict-free energy consumption using the same power model, systematic biases in absolute power largely cancel. For instance, varying drivetrain efficiency $\eta_\mathrm{drv}$ by $\pm5\%$ shifts absolute power estimates but changes relative overhead only marginally, since both numerator and denominator are affected proportionally. Relative overhead percentages are, therefore, more robust to modeling assumptions than absolute energy values. 

Second, simulations operated at cruise altitude throughout, avoiding transition phases where energy costs and modeling uncertainty are greatest. Conflicts near vertiports during these phases would likely increase both baseline energy consumption and conflict-resolution overhead, making the cruise-only results a conservative lower bound on the overall energy impact of tactical deconfliction. 

Third, the prediction model uses only pre-flight features; real-time updates during flight could improve accuracy as conflicts develop. 

Fourth, this study focused exclusively on MVP; alternative algorithms may exhibit different energy consumption characteristics.

\section{Conclusion}
\label{sec:conclusion}
This paper characterized eVTOL energy consumption under MVP-based tactical conflict resolution through large-scale simulation of approximately $71,767$ transits across traffic densities of $10-60$ aircraft in a $10$~nm sector radius. Three principal findings emerge.

First, MVP-based deconfliction is energy-efficient: median overhead remains below $1.5\%$ across all density levels, validating its suitability for energy-constrained operations. Second, the overhead distribution exhibits right-skewness with tail cases reaching $44\%$, indicating that reserve planning must account for occasional high-overhead transits; a $4-5\%$ margin accommodates $95\%$ of scenarios. Third, the machine learning model provides well-calibrated uncertainty bounds enabling targeted reserve policies that balance safety against efficiency.

Several directions merit future investigation: flight test validation of the power model, extension to full-envelope operations including climb and descent, real-time prediction updates leveraging aircraft-to-aircraft data links, and comparative analysis of alternative deconfliction algorithms for energy-sensitive operations.

Together, these results demonstrate the feasibility of decentralized tactical deconfliction for eVTOL aircraft in high-density airspace and provide quantitative guidance for reserve determination in Advanced Air Mobility operations.

\section*{Acknowledgment}
This project is supported by the NASA Grant 80NSSC23M0059 under the NASA University Leadership Initiative.

\bibliography{ref}

@techreport{faa2023uam,
  author = {{Federal Aviation Administration}},
  title = {{Urban Air Mobility Concept of Operations v2.0}},
  institution = {U.S. Department of Transportation},
  year = {2023}
}

@inproceedings{nasa2021aam,
author = {Kenneth H. Goodrich and Colin R. Theodore},
title = {{Description of the NASA Urban Air Mobility Maturity Level (UML) Scale}},
booktitle = "AIAA Scitech 2021 Forum",
doi = {10.2514/6.2021-1627},
year= {2021}
}

@book{Raymer2018,
  author    = {Raymer, Daniel P.},
  title     = {{Aircraft Design: A Conceptual Approach}},
  edition   = {6th},
  publisher = {American Institute of Aeronautics and Astronautics},
  year      = {2018},
  doi       = {10.2514/4.104909}
}

@book{shevell_fundamentals_of_flight,
    author = {Richard Shevell},
    title = {{Fundamentals of flight}},
    publisher = {Prentice-Hall} ,
    year = {1983}
}

@book{Leishman2006,
  author    = {Leishman, J. Gordon},
  title     = {{Principles of Helicopter Aerodynamics}},
  edition   = {2nd},
  publisher = {Cambridge University Press},
  year      = {2006},
  isbn      = {978-0-521-85860-1}
}

@book{Johnson2013,
  author    = {Johnson, Wayne},
  title     = {{Rotorcraft Aeromechanics}},
  series    = {Cambridge Aerospace Series},
  number    = {36},
  publisher = {Cambridge University Press},
  year      = {2013},
  isbn      = {978-1-107-02807-4}
}

@book{Hoerner1965,
  author    = {Hoerner, Sighard F.},
  title     = {{Fluid-Dynamic Drag: Practical Information on Aerodynamic Drag and Hydrodynamic Resistance}},
  edition   = {2nd},
  publisher = {Dr.-Ing. S.F. Hoerner},
  year      = {1965}
}

@inproceedings{Silva2018,
  author    = {Silva, Christopher and Johnson, Wayne and Antcliff, Kevin R. and Patterson, Michael D.},
  title     = {{VTOL} Urban Air Mobility Concept Vehicles for Technology Development},
  booktitle = {2018 Aviation Technology, Integration, and Operations Conference},
  series    = {AIAA Aviation Forum},
  year      = {2018},
  month     = jun,
  publisher = {AIAA},
  doi       = {10.2514/6.2018-3847}
}

@book{Anderson2017,
  author    = {Anderson, Jr., John D.},
  title     = {{Fundamentals of Aerodynamics}},
  edition   = {6th},
  publisher = {McGraw-Hill Education},
  year      = {2017},
  isbn      = {978-1-259-12991-9}
}

@book{anderson1999performance,
  author = {John D. Anderson},
  title = {Aircraft Performance and Design},
  publisher = {McGraw-Hill},
  year = {1999},
  isbn = {978-0-07-001971-3}
}

@inproceedings{victoria_aam_ucirvine,
author = {Victoria R. Gonzalez and Jacqueline L. Huynh},
title = {{Integrating Aircraft Performance in Traffic Flow Management Analysis for Advanced Air Mobility}},
booktitle = {AIAA AVIATION FORUM AND ASCEND 2025},
year= {2025},
doi = {10.2514/6.2025-3637}, 
}

@online{joby_msfs_pressrelease_116,
  author       = {{Joby Aviation, Inc.}},
  title        = {{Fly the Joby Aircraft in the New Release of Microsoft}},
  year         = {2024},
  url          = {https://ir.jobyaviation.com/news-events/press-releases/detail/116/fly-the-joby-aircraft-in-the-new-release-of-microsoft},
  urldate      = {2026-02-18},
  organization = {Joby Aviation Investor Relations},
  note         = {Press release}
}

@inproceedings{freeflight_hoekstra,
title = {{Free Flight in a crowded Airspace?}},
author = "JM Hoekstra and RCJ Ruigrok and {van Gent}, RNHW",
year = "2000",
language = "English",
isbn = "978-1-56347-474-3",
pages = "533--546",
editor = "GL Donohue and AG Zellweger",
booktitle = "Air Transportation \& System Engineering",
publisher = "American Institute of Aeronautics and Astronautics Inc. (AIAA)",
}

@article{HOEKSTRA_mvp,
title = {{Designing for safety: the ‘free flight’ air traffic management concept}},
journal = {Reliability Engineering \& System Safety},
volume = {75},
number = {2},
pages = {215-232},
year = {2002},
issn = {0951-8320},
doi = {https://doi.org/10.1016/S0951-8320(01)00096-5},
author = {J.M Hoekstra and R.N.H.W {van Gent} and R.C.J Ruigrok},
}

@article{eby1995self,
  title={{A self-organizational approach for resolving air traffic conflicts}},
  author={Eby, Martin S},
  journal={The Lincoln Laboratory Journal},
  volume={7},
  number={2},
  pages={239--254},
  year={1995},
  publisher={Lincoln Laboratory Lexington, MA, USA}
}

@article{hoekstra_state_based_cd_cr,
title = {{Aerial Robotics: State-based Conflict Detection and Resolution (Detect and Avoid) in High Traffic Densities and Complexities}},
author = "J.M. Hoekstra and J. Ellerbroek",
year = "2021",
doi = "10.1007/s43154-021-00061-6",
language = "English",
volume = "2",
pages = "297--307",
journal = "Current Robotics Reports",
issn = "2662-4087",
publisher = "Springer",
}

@inproceedings{Hoekstra_bluesky,
  author    = {Hoekstra, Jacco M. and Ellerbroek, Joost},
  title     = {{BlueSky ATC} Simulator Project: An Open Data and Open Source Approach},
  booktitle = {Proc. 7th Int. Conf. Research in Air Transportation (ICRAT)},
  year      = {2016},
}

@mastersthesis{Schaberg2020,
  author    = {Schaberg, W.},
  title     = {{A Decentralized Recovery Method for Air Traffic Conflicts}},
  school    = {{Delft University of Technology}},
  year      = {2020},
  address   = {Delft, The Netherlands}
}

@article{ribeiro_joost_hoekstra_review,
AUTHOR = {Ribeiro, Marta and Ellerbroek, Joost and Hoekstra, Jacco},
TITLE = {{Review of Conflict Resolution Methods for Manned and Unmanned Aviation}},
JOURNAL = {Aerospace},
VOLUME = {7},
YEAR = {2020},
NUMBER = {6},
ARTICLE-NUMBER = {79},
ISSN = {2226-4310},
DOI = {10.3390/aerospace7060079}
}

@article{xuxi_uam_nyc,
author = {Yang, Xuxi and Wei, Peng},
year = {2020},
month = {05},
pages = {1-14},
title = {{Scalable Multi-Agent Computational Guidance with Separation Assurance for Autonomous Urban Air Mobility}},
volume = {43},
journal = {Journal of Guidance, Control, and Dynamics},
doi = {10.2514/1.G005000}
}

@techreport{nasa2025tactical,
  author      ={Yarramreddy, Gautam and de Alvear Cardenas, Jose Ignacio and Pradeep, Priyank and Xue, Min and Lee, Seungman and Kuo, Vincent},
  title       = {{Simulation Framework for Tactical Separation Assurance}},
  institution = {NASA},
  number      = {NASA-TM-20250002761},
  year        = {2025}
}

@techreport{nasa2020tactical,
  author = {Johnson, M. and Larrow, J.},
  title       = {{UAS traffic management conflict management model}},
  institution = {NASA},
  number      = {NASA/TM-2020-5002076},
  year        = {2020}
}

@article{sripad_venkatasubramanian_energy_efficient_battery,
author = {Shashank Sripad  and Venkatasubramanian Viswanathan },
title = {{The promise of energy-efficient battery-powered urban aircraft}},
journal = {Proceedings of the National Academy of Sciences},
volume = {118},
number = {45},
pages = {e2111164118},
year = {2021},
doi = {10.1073/pnas.2111164118},
}

@article{DAI_modeling_power_consumption_est_quad,
title = {{Data-efficient modeling for power consumption estimation of quadrotor operations using ensemble learning}},
journal = {Aerospace Science and Technology},
volume = {144},
pages = {108791},
year = {2024},
issn = {1270-9638},
doi = {https://doi.org/10.1016/j.ast.2023.108791},
author = {Wei Dai and Mingcheng Zhang and Kin Huat Low},
}

@inproceedings{ayalew2023data,
  title={{Data-driven urban air mobility flight energy consumption prediction and risk assessment}},
  author={Ayalew, Yonas and Bedada, Wendwosen and Homaifar, Abdollah and Freeman, Kenneth},
  booktitle={Intelligent Systems Conference},
  pages={354--370},
  year={2023},
  organization={Springer}
}

@article{Pelegrn2023UrbanAM,
  title={{Urban air mobility: from complex tactical conflict resolution to network design and fairness insights}},
  author={Mercedes Pelegr{\'i}n and Claudia D’Ambrosio and R{\'e}mi Delmas and Youssef Hamadi},
  journal={Optimization Methods and Software},
  year={2023},
  volume={38},
  pages={1311 - 1343},
}

@ARTICLE{chen_dcb_deconfliction,
  author={Chen, Shulu and Evans, Antony D. and Brittain, Marc and Wei, Peng},
  journal={IEEE Transactions on Intelligent Transportation Systems}, 
  title={{Integrated Conflict Management for UAM With Strategic Demand Capacity Balancing and Learning-Based Tactical Deconfliction}}, 
  year={2024},
  volume={25},
  number={8},
  pages={10049-10061},
  doi={10.1109/TITS.2024.3351049}
  }

@misc{brittain2019autonomousairtrafficcontroller,
      title={{Autonomous Air Traffic Controller: A Deep Multi-Agent Reinforcement Learning Approach}}, 
      author={Marc Brittain and Peng Wei},
      year={2019},
      eprint={1905.01303},
      archivePrefix={arXiv},
      primaryClass={cs.LG}, 
      url={https://arxiv.org/abs/1905.01303}
}

@INPROCEEDINGS{brittain_autono_separation_assurance,
  author={Brittain, Marc and Wei, Peng},
  booktitle={2019 IEEE Intelligent Transportation Systems Conference (ITSC)}, 
  title={{Autonomous Separation Assurance in An High-Density En Route Sector: A Deep Multi-Agent Reinforcement Learning Approach}}, 
  year={2019},
  volume={},
  number={},
  pages={3256-3262},
  doi={10.1109/ITSC.2019.8917217}}

@article{taye2025trajectory,
  author    = {Taye, Abenezer G. and Wei, Peng},
  title     = {{Energy-Efficient Trajectory Planning and Feasibility Assessment Framework for Drone Package Delivery}},
  journal   = {{AIAA Journal of Aerospace Information Systems}},
  year      = {2025},
}

@inproceedings{taye2024feasibility,
  author    = {Taye, Abenezer G. and Wei, Peng},
  title     = {{Flight Mission Feasibility Assessment of Urban Air Mobility Operations under Battery Energy Constraint}},
  booktitle = {{AIAA SciTech Forum}},
  year      = {2024},
}

@inproceedings{thompson2023octocopter,
  author    = {Thompson, Ellis L. and Taye, Abenezer and Ashby, James and Fattah, Gerald and Wei, Peng and Bonin, Thomas and Jones, James C. and Quinones-Grueiro, Marcos and Biswas, Gautam},
  title     = {{Probabilistic Evaluation for Flight Mission Feasibility of a Small Octocopter in the Presence of Wind}},
  booktitle = {{AIAA Aviation Forum}},
  year      = {2023},
}

@inproceedings{taye2025strategic,
  author    = {Taye, Abenezer and Chen, Shulu and Wei, Peng},
  title     = {{Energy-Aware Strategic Traffic Management for Urban Air Mobility}},
  booktitle = {{AIAA SciTech Forum}},
  year      = {2025}
}

@misc{aziz2026transformerbasedmultiagentreinforcementlearning,
      title={{Transformer-based Multi-agent Reinforcement Learning for Separation Assurance in Structured and Unstructured Airspaces}}, 
      author={Arsyi Aziz and Peng Wei},
      year={2026},
      eprint={2601.04401},
      archivePrefix={arXiv},
      primaryClass={cs.RO},
}
\bibliographystyle{IEEEtran}

\end{document}